\documentstyle[twocolumn,prl,aps,psfig]{revtex}

\newcommand{\half}{\frac 1 2 }
\newcommand{\eg}{{\em e.g.} }
\newcommand{\ie}{{\em i.e.} }
\newcommand{\be}{\begin{eqnarray}}
\newcommand{\ee}{\end{eqnarray}}

\newcommand{\etal}{{\em et al.\ }}

\newcommand{\pref}[1]{(\ref{#1})}

 \def\phidag{\phi^{\dagger}}
 \def\fii{\varphi}
 
 \def\chidag{\chi^{\dagger}}

\def\aeo{a_1^0}
\def\ato{a_2^0}
\def\aev{\vec a_1}
\def\atv{\vec a_2}
\def\sz{\, \vec\sigma\cdot \hat m}

\def\nh{\hat n}

\def\at{\tilde a}
\def\atvidv{\vec{\tilde a}}
\newcommand{\ki}[2]{l_{#1 #2}}

\def\rhobar{\bar\rho}

\def\mh{\hat m}
\def\rup{\rho_{\uparrow}}
\def\rdown{\rho_{\downarrow}}

\begin{document}

\draft
\preprint{USITP-98-17}

\title{A Field Theory for Partially Polarized Quantum Hall States}

\author{T.H. Hansson$^{1}$, A. Karlhede$^{1}$, J.M.
Leinaas$^2$ and U. Nilsson$^3$}

\address{$^1$ Department of Physics, Stockholm University, Box 6730,
S-11385 Stockholm, Sweden}
\address{$^2$ Deptartment of Physics, University of Oslo, P.O. Box 1048 Blindern,
N-0316 Oslo, Norway}
\address{$^3$ Department of Applied Mathematics, University of Waterloo, Ontario, Canada,
N2L 3G1}

\date{\today}
\maketitle

\begin{abstract}
We propose a new  effective field theory for partially polarized quantum Hall
states. The density and polarization for the mean field ground states are
determined by couplings to two  Chern-Simons gauge fields.  In addition there
is  a $\sigma$-model field, $\mh$, which is necessary both to preserve  the
Chern-Simons gauge symmetry that determines the correlations in the ground
state, and the global SU(2) invariance related to spin rotations.  For states
with non zero polarization, the low energy dynamics is that of a ferromagnet.
In addition to  spin waves, the spectrum contains topological solitons, or
skyrmions, just as in the fully polarized case. The  electric charge of the
skyrmions is given by $Q_{el}=\nu P Q_{top}$, where $\nu$ is the filling
fraction, $P$ the magnitude of the polarization, and $Q_{top}$ the
topological charge. For the special case of full polarization, the theory
involves a single scalar field and a single Chern-Simons field in addition to
the $\sigma$-model field, $\mh$. We also give a heuristic derivation of the
model lagrangians for both full and partial polarization, and show that in a
mean field picture,   the field $\mh$ is necessary in order to take into
account the  Berry phases originating from  rotations of the electron spins
\end{abstract}

\pacs{73.40Hm}

\section{Introduction}

\renewcommand{\theequation}{1.\arabic{equation}}
\setcounter{equation}{0}
\vskip 5mm
It is known that at low electron densities fractional quantum Hall (QH)
ground states at certain fractions such as $2/5$ and $4/3$ are not fully
polarized. This has been established by studying the transport properties at
a fixed filling fraction as a function of the applied magnetic field $\vec B$
(either by tilting the field or by changing the electron
density)\cite{clark,eisenstein,furneaux}. Plateaux that are destroyed by
increasing the magnetic field, \ie by increasing the Zeeman gap, are natural
candidates for fractional QH states with non-maximal polarization, and
observed crossover behavior is consistent with transitions from unpolarized
to partially or fully polarized states. This picture is supported by
numerical calculations (exact diagonalization of few electron systems) and
theoretical considerations based on Halperin wave-functions and hierarchical
schemes\cite{chakraborty,girvin2}.

In another line of development progress has been made in understanding
spin effects in fully
polarized systems.  Sondhi \etal  predicted that the lowest energy charged
excitations are skyrmions\cite{skkr}. This was then experimentally
confirmed by measuring the excess spin of these
quasiparticles\cite{barrett,schmeller,goldberg,eaves}.

An important tool in studying skyrmions (and spin textures in general)
is  effective theories for the spin degrees of freedom. For the
simplest case of a fully polarized state, the effective theory is
a non-linear $\sigma$-model given by,
\be
{\cal L}_{eff} &=& \bar\rho{\cal L}_{kin}
- \frac {\kappa\bar\rho} 4 (\vec \nabla\hat m)^2  \nonumber \\
&-& V(\bar\rho + \delta\rho) + \half \mu_{e} \bar\rho \vec B\cdot
\hat m  \
\ \ \ \ .  \label{sig}
\ee
Here $\vec B$ is the magnetic field, $V(\delta \rho)$ is the effective
Coulomb potential and
${\cal L}_{kin}$ is the kinetic term for a ferromagnet, defined by its
variation,
$\delta {\cal L}_{kin}/\delta\hat m^i =
\epsilon^{ijk}\hat m^j\partial_0\hat m^k$.\footnote{
${\cal L}_{kin}$
cannot be written as a local function of $\hat m$. It has the same form as
the action for a charged particle on a sphere moving in
the field of a unit magnetic monopole.}
The unit vector $\hat m$ describes the magnetization and
the deviation $\delta \rho$ from the ground state charge density $\bar\rho$ is
proportional to the topological (Pontryagin) charge  density $q=\mh \cdot (
\partial_x \mh \times \partial_y \mh )/4\pi$: $\delta\rho = \nu q$, where
$\nu$ is the filling
fraction. In the case
of a single filled Landau level, $\nu=1$, the
effective lagrangian (\ref{sig}) can be derived from first principles
\cite{indiana,ray,ray2}.

In this paper we present an effective theory that has partially
polarized ground states; the excitations corresponding to fluctuations
in the density and the magnitude of the polarization both have large
gaps; the low-energy excitations are ferromagnetic spin waves with the
expected Zeeman gap; the effective low energy lagrangian is again a
$\sigma$-model, which in addition to spin waves
also has skyrmion solutions; in the limit of zero polarization
the spin waves decouple, and the electric charge of the skyrmions vanishes.
Our model is a natural, and in a sense minimal, extension of the
Chern-Simons-Landau-Ginzburg model described in  ref. 14. That
model had partially polarized ground states, but
did  not account correctly for the low-energy excitations, whereas the present
model does. The limiting case of full polarization is treated
separately, and the effective low energy lagrangian is again a
$\sigma$-model.

We do not have a firm microscopic justification of our model
lagrangian, but we will provide  a heuristic derivation
which we believe captures  important pieces of the  physics.

The paper is organized as follows. In the next section we present our
model lagrangian and discuss  its properties. In section 3 we derive the
effective low energy lagrangian which is of
sigma model type, and derive the spin wave spectrum. Section 4
deals with the skyrmion solutions, and section 5 with the microscopic
derivation of the model lagrangian with a special discussion of the case
of full polarization. We conclude in section 6 with some
comments about the status of the microscopic derivation, and some
general remarks.

\section{The model lagrangian }

\renewcommand{\theequation}{2.\arabic{equation}}
\setcounter{equation}{0}
\vskip2mm
Consider a two dimensional electron gas subject to a magnetic field of
constant magnitude and direction, $\vec B = \nabla \times \vec A$. The
magnetic field is in general tilted relative to a normal vector of the
electron plane. We model this system with the following lagrangian (density),
\be
{\cal L} &=& \phidag iD_0 \phi
 -\frac 1 {2m_{e}} |\vec D\phi|^2
   - \frac 1 {2\pi} \ki \alpha \beta
   \epsilon_{\mu\nu\sigma}a_\alpha^\mu\partial^\nu a_\beta^\sigma
\label{lag} \nonumber \\
 &-& V(\rup , \rdown)  - \frac {V_0} 2 (\partial _i \vec S)^2
 + \mu_{e}  \vec S \cdot \vec B \ \ \ ,
\ee
where the covariant derivatives  $D_0$ and $\vec D$ are given
by\footnote{
The three-vector notation is only for notational convenience; the metric
is Euclidian, \ie $D^{\mu} = D_{\mu}$, and all signs are written
explicitly.}
\be
iD^\mu &=& i\partial^\mu + a_1^\mu + a_2^\mu\sz + A^\mu  -\half
(\mh\times\partial^\mu \mh)\cdot\vec\sigma \ , \label{cov}
\ee
and the spin density $\vec S$ is related to the unit vector field
$\mh$ by  $\vec S = (1/2) \,\mh \, \phidag
\vec \sigma \cdot \mh \,\phi$, as we shall discuss below.
The densities $\rup$ and $\rdown$ are defined with respect to the
quantization direction $\mh$: $\rho (\mh \cdot \nh) = \phidag
\mh\cdot  \vec \sigma \phi =
\rup - \rdown$,  where  $\rho = \phidag\phi=\rup + \rdown$ and
$\rho\hat n = \phidag\vec\sigma\phi$.

The field content
differs from that in the conventional Landau-Ginzburg description
\cite{wen,ezawa} in that in addition to the two-component complex {\em
bosonic} field $\phi$, and the Chern-Simons (CS) fields
$a_\alpha^\mu$, there is also a vector field $\hat m$
describing the direction of the spin polarization.
We use a notation where $i,j$... $= 1,2$; $\mu$,
$\nu$....$=0,1,2$; $m_{e}$ is the electron mass; $\mu_{e}$ is the effective
magnetic moment of the electron, $V_0$ is an interaction parameter and $\hbar
= e = c =1$. The elements of the symmetric matrix
$l^{-1}$ are integers whose diagonal elements are both either even
or odd.\footnote
{The generalized Halperin $(m_{1},m_{2},n)$ states correspond to the
following values for $l_{\alpha\beta}$:
$\Delta l_{11}=m_{1}+m_{2}-2n$, $\Delta l_{22}=m_{1}+m_{2}+2n$
and $\Delta l_{12}=m_{2}-m_{1}$, where $\Delta =
m_{1}m_{2}-n^{2}$. }

It is convenient to decompose $\phi$ as  $\phi = \fii \chi$,
where  $\chidag\chi=1$,
$\fii  = \sqrt{\rho}$, and the CP(1) field $\chi$
is related to the  vector $\nh$ by $\nh = \chidag\vec\sigma \chi$.
The two gauge potentials
$a^\mu_1$ and $a^\mu_2$  couple to the charge and the $\hat m$-component of the
density $\rho\hat n$ respectively.  The degrees of freedom in $\chi$
can conveniently be thought of as the two angles
describing the direction of $\nh$ plus an additional  overall phase,
$e^{i\theta(x)}$.
(Note that it is  $\mh$ and not $\hat n$ that is  identified with the
local direction of polarization.)

We now discuss the symmetry properties of (\ref{lag}) by considering
the following transformations,
\be
\chi \rightarrow e^{i\alpha(x)}\chi \ \ , \ \ \ a_{1}^{\mu}
&\rightarrow& a_{1}^{\mu} + \partial^{\mu}\alpha(x)  \label{gauge1} \\
\chi \rightarrow e^{i\beta(x)\vec \sigma \cdot \mh (x)}\chi \ \ , \ \ \
a_{2}^{\mu} &\rightarrow&
a_{2}^{\mu} + \partial^{\mu}\beta(x) \label{gauge2} \\
\chi \rightarrow  e^{\frac i 2 \vec k\cdot\vec\sigma}\chi  \ \ , \ \ \ \mh
&\rightarrow&
e^{i\vec k\cdot \vec L}\mh \label{global} \ \ \ ,
\ee
where the $3 \times 3$ matrix $\vec L$ is the  angular
momentum in the vector representation.
While $\rho$ is invariant under the two U(1) gauge transformations
\pref{gauge1}
and \pref{gauge2}, the unit vector $\nh$ is only invariant under
the  one  related to $a_1$, while under the one related to $a_2$,
it rotates around the $\hat m$-axis as $\nh \rightarrow e^{2i\beta(x)\vec
L\cdot \hat m}\nh$.
From this follows that both the densities, $\rup$ and $\rdown$,
and the polarization, $P=\mh\cdot\hat n=\cos\alpha$,
are invariant under the two gauge transformations and,
consequently, so are also the (in general non-local) potential $
V(\rup , \rdown) $ and the Zeeman term in \pref{lag}.
A simple calculation shows that the covariant derivative \pref{cov}
transforms homogeneously under both gauge transformations, and thus the
full lagrangian \pref{lag} is gauge invariant.

In the limit of vanishing Zeeman coupling, \ie $\mu_e = 0$, the
lagrangian is also invariant under the {\em global} SU(2) symmetry
\pref{global}, corresponding to  simultaneous constant
rotations of  $\phi$ and $\mh$. Using the SU(2)  invariant part of the action
alone we can use the  Noether procedure to derive the generator
$\vec{\cal S}$, which is the integral of the spin density $\vec S$,\footnote
{Note that
$\vec S$ gets contributions from variations both in $\phi$ and $\mh$.}
\be
\vec {\cal S} = \int d^{2}x\,  \vec S = \half \int d^{2}x\,
\rho (\hat m\cdot \hat n) \hat m \label{pol} \ \ \ \ \ .
\ee
From this it follows that  $\mh$ should be identified with the direction,
and $\mh\cdot\hat n$  with the magnitude, of the polarization.

A homogeneous ground state of the model (\ref{lag}) can be found in the
following way.  We choose Coulomb gauge ($\vec\nabla\cdot\vec a_{\alpha}^i =0$)
and  look for a solution which minimizes each of the terms of the
Hamiltonian separately.
The Zeeman energy and the gradient energy
$\sim (\nabla
\hat S)^2$ are minimized by taking $\phi$ as an arbitrary constant,
and  $\mh$ equal to a constant unit vector
$\mh^{0}$  pointing in the direction of $\vec B$.
In the basis where $\vec\sigma\cdot\hat m$ is
diagonal, the two first terms in (\ref{lag}) are  diagonal and describe
two scalar fields $\phi_-$ and $\phi_+$ coupled to the statistical vector
potentials
$\vec a_- = \aev - \atv$ and $\vec a_+ = \aev + \atv$
respectively. The kinetic energy is minimized by $\phi =
const.$, and $\vec a_+ = \vec a_- = -\vec A $. Varying $\cal L$ with
respect to $a^0_{\alpha}$ gives the constraints  $\pi\rho = -\ki
1 1 b_1 - \ki 1 2 b_2 $ and  $\pi\rho (\hat m\cdot\hat n)= -\ki
1 2 b_1 - \ki 2 2 b_2 $.
The solution corresponding to minimal kinetic energy therefore is
\be
\rho &=& \ki 1 1 B_\perp /\pi \equiv \rhobar \nonumber \\
P &=&  \hat n\cdot \hat m =\cos\alpha = {\ki 1 2 }/{\ki 1 1 }\equiv \bar P
\; ,
\label{gnst}
\ee
where $B_\perp$ is the component of $\vec B$ perpendicular to the plane.
In order for the density $\rhobar$ to correspond to minimal potential
energy, a chemical potential has to be included in the lagrangian.
The value of the chemical potential is then fixed by this requirement.
Similarly, the requirement that the polarization $\bar P$ should correspond to
minimal potential energy will fix the value of the Zeeman term. However, one
should note that the ground state \pref{gnst} is supposed to exist for a
range of values of the chemical potential and the Zeeman energy. This is
because a change in these quantities can be absorbed in a change in
$a_+^0$ and $a_-^0$. Such a change will increase the energy, but not change the
ground state until the gap of the excitation energy is exceeded. As in
ref. 14,
we interpret this mean field ground state as a quantum Hall state with filling
fraction
$\nu=2\pi \bar \rho /B_\perp = 2{\ki 1 1} $, and polarization
$\bar P $.

\section{The effective $\sigma$-model}

\renewcommand{\theequation}{3.\arabic{equation}}
\setcounter{equation}{0}
\vskip2mm

\newcommand{\ab}{\bar\alpha}
\newcommand{\ca}{\cos\alpha}
\newcommand{\sa}{\sin\alpha}
\newcommand{\tho}{\vartheta}
\newcommand{\tht}{\theta}
\newcommand{\eone}{\hat e_1}
\newcommand{\etwoz}{\hat e_2^0}
\newcommand{\eonez}{\hat e_1^0}
\newcommand{\etwo}{\hat e_2}
\newcommand{\mhz} {{\hat m^0}}

Since the mean field ground state given above spontaneously breaks
the approximate SU(2) symmetry of the model, we expect Goldstone modes with a
gap given by the Zeeman energy $\mu_e B$.
These modes are spin waves  where both $\hat m$ and $\hat n$ vary, but the total
density, $\rho$, and the magnitude of the polarization, $\mh\cdot\hat n$,
remain fixed. To find these modes, we parametrize $\phi$ as:
\be
\phi=\sqrt\rho e^{i\tho }e^{i\frac \alpha 2 \vec\sigma\cdot\eone(\tht)}
\chi_{\mh} \ \ \ \ \ , \label{par}
\ee
where the  spinor $\chi_{\mh}$ is choosen so that
\be
\chidag_{\mh} \vec\sigma \chi_{\mh} = \mh \ \ \ \ \ ,
\ee
and we have introduced an orthonormal basis $(\mh,\, \eone,\, \etwo)$.
The four degrees of freedom in the complex two-spinor field
$\phi$ are coded in the density $\rho$, and the three angles $\alpha$,
$\tho$ and $\tht$. Since $\alpha$ is measured relative to the vector
$\mh$  ($\cos \alpha = \mh\cdot \nh$) we expect high frequency modes
for the  fluctuations in $\alpha$ as well as in
$\rho$. $\tht$ parametrizes rotations of $\nh$ around $\mh$ and can be removed
by an $a_2$ gauge transformation \pref{gauge2}; similarly, $\tho$ can be
removed
by an $a_1$
gauge transformation \pref{gauge1}.
Next we  introduce the constant  basis  $(\mhz,\,
\eonez,\, \etwoz)$, and taking $\mhz = \hat z$ we have the following
explicit parametrization,
\be
\mh = R(\vec k)\mhz  &=&  (\sin k\cos\beta,\sin k\sin\beta,\cos k)
\label{mh} \\
\eonez = \hat k &=& (-\sin\beta, \cos\beta,0)
\ee
and (with $\etwoz = \mh\times\hat k $),
\be
\eone &=& \cos\tht\eonez + \sin\tht\etwoz \ \ \ \ \ .
\ee
We also define the topological vector potential
\be
\tilde a^\mu = \chidag_{\mh} i\partial^\mu \chi_{\mh} =
\sin^2\frac k 2 \partial^\mu\beta  \label{atilde} \ \ \ \ \ .
\ee
It is now only a matter of  algebra to rewrite
\pref{lag} in the following form,
\be
{\cal L} &=& \rho[\aeo + \ca\ato + \ca\tilde a^0 ] \nonumber \\
&-&\frac {\rho} {8m_{e}} [\frac 1 \rho (\vec\nabla\rho)^2 +
(\vec\nabla\alpha)^2  ]  -
V(\rup,\rdown)   \nonumber \\
&-&\frac \rho {2m_{e}} \{[\aev + \vec A + \ca (\atv + \atvidv) ]^2
+ \sin^{2}\alpha (\atv + \atvidv )^2 \} \nonumber \\
&-&\frac 1 {2\pi} \ki \alpha \beta
\epsilon_{\mu\nu\sigma}a_\alpha^\mu\partial^\nu
a_\beta^\sigma
-\frac {V_0} 2 (\partial_i \vec S)^2  +
\mu_{e}\vec B\cdot\vec S \ \ ,   \label{efflag}
\ee
where we  fixed a unitary gauge by $\tho=0$ and $\eone(\tht) = \hat k$
(the fields $\tho$ and $\tht$ were absorbed in the longitudinal parts
of $a_1$ and $a_2$ respectively), and also made the variable changes
$a_{1}^{\mu} - \half \partial^{\mu}\beta \rightarrow a_{1}^{\mu}$
and $a_{2}^{\mu} + \half \partial^{\mu}\beta \rightarrow
a_{2}^{\mu}$.

It is easily verified that the  topological vector potential $\tilde
a^\mu$ is related to the
ferromagnetic kinetic term in \pref{sig}, and the Pontryagin
density, $q$, by,
\be
{\cal L}_{kin} &=& \tilde a^{0}  \\
2\pi q &=&  \tilde b = \vec \nabla  \label{pontryagin}
\times \vec {\tilde  a}
\ \ \ \ \ .
\ee
Note that in the parametrization \pref{mh}, ${\cal L}_{kin}$ becomes local
at the expense of introducing an arbitrary fixed direction
$\mh^0$.

To study the fluctuations around the mean field solution, we decompose the gauge
fields into transverse and longitudinal parts,
\be
\vec a_\alpha = \vec a^T_\alpha + \vec\nabla\theta_\alpha\ \ \ \ \ ,
\ee
use the constraints to express the transverse components in terms of
densities, and expand \pref{efflag} to
quadratic order in the small parameters $\theta_\alpha$, $\delta\rho$
and $\delta\alpha$. Note that since the lagrangian is first order in
time derivatives, these four variables describe only two
independent  modes corresponding to the fluctuations
in the density and the magnitude of the polarization.
As expected, these modes are precisely those of the model with fixed $\mh$ and
have gaps given by,
\be
\omega_\rho &=& \omega_c \label{modes}  \\
\omega_\alpha &=& 2\pi (l^{-1})_{22}\nu\sin^2\alpha\,
         \omega_c \ \ \ \ \ , \label{almode}
\ee
where $\omega_c=B_\perp/m_e$ is the cyclotron frequency. Note that full
polarization ($\alpha=0$) is a special case, which will be
treated separately in section 5 . In the two-component formulation it corresponds to a
non-invertible $l$-matrix and the second mode $\omega_\alpha$ is not
present.

Since the modes \pref{modes} and \pref{almode} have large gaps, the corresponding (fast)
variables $\theta_\alpha$, $\delta\rho$
and $\delta\alpha$ can be integrated out to give an effective
lagrangian in the low-energy variable $\mh$. As a lowest order approximation
this can be done by fixing the fast variables through the requirement of
minimal energy in a general background field $\mh (x)$. The corresponding
ground state is, for a slowly varying field $\mh (x)$, characterized
by vanishing fields $b_1+B_\perp=0$ and $b_2+\tilde b=0$.  In a general
background this implies that the  charge and spin densities are not
constant, but are related to the topological density $q$ of the
background field in the following way,
\be
\rho &=& \bar\rho + \frac{l_{12}} \pi \tilde b = \bar \rho +
\nu  \cos\ab\, q \nonumber \\
\rho\cos\alpha &=& \bar\rho\cos\ab +
  \frac {l_{22}} \pi \tilde b  =  \bar\rho\cos\ab +2{l_{22}} q \ \ \ . \label{densities}
\ee
With these expressions inserted in the lagrangian one obtains for the
effective lagrangian of the low-energy variable $\mh$,
\be
{\cal L}_\sigma &=& \rho \cos\alpha \,{\cal L}_{kin}
-\frac {V_0} 2 (\rho \cos\alpha)^{2} (\vec\nabla\mh)^2  \nonumber \\
&+&\frac {\mu_{e}} 2 \cos\alpha\, \vec B\cdot\mh - V(\rup,\rdown)  \ \ \
,\label{eff}
\ee
where $\rho$ and $\rup-\rdown=\rho \cos \alpha$ are determined by $q(\mh)$
via \pref{densities}. (Derivative terms in $q$ have here been
neglected.) Higher order terms in the loop expansion
give derivative corrections to this result. From
\pref{eff} we obtain the  following spin wave dispersion relation to second
order in the momentum $p$,
\be
\omega = \mu_e B + \frac \kappa {\cos\ab} p^2 \; , \label{disp}
\ee
with $\kappa=(V_0/2)\rho\cos^2\ab$. The effective lagrangian \pref{eff}
is valid for full as well as for partial polarization. Note that for
$\alpha=0$, \ie the fully polarized case, we retain the result of
Sondhi \etal\cite{skkr} with correct normalization of the gap. For
$\alpha=\pi/2$ the kinetic term in (\ref{eff}) vanishes and there is no
propagating spin wave.

\section{Skyrmion solutions }

\renewcommand{\theequation}{4.\arabic{equation}}
\setcounter{equation}{0}
\vskip2mm
The low energy effective lagrangian \pref{eff} gives rise
to the following hamiltonian,
\be
{\cal H} =  \frac {\kappa\rho} 4 (\vec\nabla\mh)^2 - \frac{\mu_e} 2
\rho\cos\alpha\, \vec B\cdot\mh + V(\rup,\rdown) \; , \label{sig2}
\ee
We recognize the
hamiltonian of the usual sigma model, and conclude that -- in the limit
of vanishing Zeeman and Coulomb
interactions -- there are scale invariant skyrmion solutions. Just as in
the fully polarized case, described by \pref{sig}, the scale is set by
a competition between Zeeman and Coulomb interactions.

According to Eq. \ref{densities}, the deviations in both the charge density,
$\delta\rho$, and in
the spin density in the $\mh$ direction, $\delta(\vec S\cdot \mh)$, are proportional
to the Pontryagin density, $q$. Let us first discuss the
consequences of the charge relation:
A skyrmion with topological charge $Q_{top}$ has
electric charge $Q_{el}$,
\be
Q_{el} = \nu  \cos\ab\, Q_{top} \; ,
\ee
so for fully polarized states, this relation is the same  as found by
Sondhi {\it et al.}, but in general the charge is proportional to the
polarization, $\bar P = \cos\ab$. In our case there is
also another  conserved quantity, namely the  total spin  in the $\mh$ direction
corresponding to the integral of the second relation in
\pref{densities}, $ \int d^2x  \,
\delta(\vec S\cdot \mh) =
l_{22} Q_{top}$. That this quantity is quantized (and in general small) does
not imply that the spin is small since
$\mh$ varies, and the spin is given by the Noether charge \pref{pol}
which generates the global SU(2) symmetry as discussed in
section 2. Also note that, since
$\vec S \sim P \mh$, the ratio of charge to spin depends only on the
skyrmion profile, not on $P$.

\section{Microscopic considerations }

\renewcommand{\theequation}{5.\arabic{equation}}
\setcounter{equation}{0}
\vskip2mm
In this section we shall give a heuristic derivation of \pref{lag} for
the special case of a local potential and vanishing Zeeman coupling.
The basic idea is to rewrite the functional integral
using auxiliary fields in such a way that the expected ferromagnetic
Goldstone mode is explicitly exhibited. Only when the theory is formulated
in such
a manner can we expect to correctly capture the long wave length physics
in a mean field approximation. In the case of spin excitations
in the Hubbard model, this was emphazised by Shultz\cite{schultz}, and in
the present context by Moon \etal\cite{indiana}. In those papers it was
also stressed that in order to reproduce  Hartree-Fock results, one
must use a particular form for the auxiliary field action. We shall see
that also in our case the detailed result depends on the particular
mean field decomposition. However, we want to stress from the outset
that the general form \pref{lag} of the lagrangian is a generic
result.

The initial steps are the same as in the
derivation of the effective sigma model \pref{sig}
for a single, filled and fully polarized
Landau level as given by Moon \etal\cite{indiana}.
In particular, we   use a local repulsive potential,
which is SU(2) invariant, and can be re-expressed in terms of spin variables:
\be
V&=&\frac {V_{0}} 2 \rho(x)^{2}= V_{0}\, \rup\rdown(x) \nonumber \\
&=& -\frac{V_{0}} 2 \,  \vec
s\cdot\vec s(x) + \frac {V_{0}} 8 \rho(x)^{2} \; . \label{decomp}
\ee
Here,  $\vec s$ is the spin operator
$\vec s =\half \psi^{\dagger}\vec\sigma\psi$, and $\rho =\psi^{\dagger} \psi$,
where $\psi$ is a two-component fermion field.\footnote{
The first equality in \pref{decomp} holds only in a functional
formulation, where $\psi$ and $\psi^{\dagger}$ are Grassman numbers.
If they are fermion operators, there are contact terms linear in the
density that can be absorbed in the chemical potential.}
This particular decomposition  differs from the one used by Moon {\it et
al.}, and we will
comment on this in section 6.
Introducing the Hubbard-Stratonovich fields $\vec h$ and $\chi$,
the euclidean partition function can  be written as,
\be
Z[A^{\mu},T] = \int {\cal D}\vec h {\cal D}\chi
e^{ -\int_{0}^{1/T} d\tau \int d^{2}x\, (\frac 1 {2V_{0}}h^{2} + \frac
2 {V_0} \chi^{2} ) }  \label{part} \nonumber \\
\times {\rm Tr}
\left\{ e^{-\frac 1 T \hat H_{0}} {\rm T}_{\tau}e^{-
\int_{0}^{1 /T} d\tau \int d^{2}x\, \left[-\vec h\cdot\vec s
-(i\chi-\mu)\psi^{\dagger}
\psi \right] }  \right\} \ \  ,
\ee
where the (2nd quantized)  hamiltonian operator $\hat H_{0}$ is given by,
\be
\hat H_{0} = \int d^{2}x\, \psi^{\dagger}\frac 1 {2m}\left[ \vec p -  \vec
A(\vec x)\right]^{2}\psi(\vec x)  \; .
\ee
Here, $T$ is the temperature
and $\mu$ is the chemical potential.
The trace in \pref{part} is taken over all anti-periodic solutions to
the following many-body problem: $N$ non-interacting spin half fermions
moving in  two dimensions in an external (2-dimensional) gauge potential $\vec
A(\vec x)$ where the spins are coupled to an external (3-dimensional)
magnetic field $\vec h = h\mh$. Note that the trace is taken over
all $N$ corresponding to a grand canonical ensemble.

The external field problem defined by the trace in \pref{part} is in
general  very
hard to solve.  To proceed we  make the
simplifying assumption that the spin of the particles in the $\mh$-direction
is a good quantum number which can be used to label the
particles. This ``adiabatic'' assumption means that we neglect the
spin-flip transitions induced by the Zeeman-like interaction $\vec
h\cdot\vec s$,\footnote
{Note that we here neglect the effect of the real
Zeeman coupling to the real external magnetic field $\vec B = \nabla \times
\vec A$.}
and below we argue that the dynamics of our
system is such that this is a  good approximation for the
low-energy sector. To evaluate the trace in
\pref{part} we use a first quantized path integral,
where the effect of the spins of the  particles can  approximately
be taken into account by including the appropriate Berry phases and Zeeman
energies,
\be
{\rm Tr} \left\{ e^{-\frac 1 T \hat H_{0}} {\rm T}_{\tau}e^{-
\int_{0}^{1/T} d\tau \int d^{2}x\, \left[-\vec h\cdot\vec s
-(i\chi-\mu)\psi^{\dagger}
\psi \right]}  \right\} \nonumber \\
=\sum_{N=0}^{\infty}\int_{\Gamma_{i}}\prod_{i=1}^{N}{\cal D}\vec
x_{i}(\tau) e^{-S_{E}[\vec x_{i}]} \  .\label{trace}
\ee
Here the sum over classical paths is understood to include sign
factors appropriate to Fermi statistics, and the euclidean action is given by
\be
S_{E} &=& \sum_{i=1}^N   [  \int_{0}^{1/T} d\tau\,
    (  \frac m {2} \dot {\vec x_{i}}^{2} -i \vec A(\vec x_{i})\cdot
\dot{\vec x_{i}}  + \mu
\nonumber \\
&-&i\chi(\vec x_{i})  + E_{M}(\vec x_i) )      - i\gamma[\Gamma_{i}] ]
   \ \ ,
\label{action}
\ee
where $\gamma[\Gamma_{i}]$ is the  Berry phase picked up by  particle
$i$ when it moves in the field $\vec h$ along the path $\Gamma_{i}$,
and $E_{M}(\vec x_i)$ is the corresponding magnetization energy.
Now comes the main observation: Using the parametrization
\pref{mh}, with the identification $\vec h = h \hat m$,  the Berry phase
$\gamma[\Gamma_{i}]$
can be written in the following way\footnote{
The Berry phase entering in \pref{action} is nothing
but the noncovariant form of the spinfactor,
as discussed in \eg ref.\cite{ghk}. In this reference it is also shown how
to express the spinfactor in terms of a Grassmann integral. It would be
interesting to formulate the present problem in this language and
retain \pref{action}, after integrating over the appropriate Grassmann
variables.       }
\be
\gamma[\Gamma_{i}] =&\mp& \oint d\beta\, \sin^{2}\frac {k(\beta(\tau)) }
2 \nonumber \\
   =&\mp&  \int_{0}^{1/T} d\tau \, \partial_{\tau}\beta \sin^{2}\frac
{k(\beta(\tau)) } 2 \nonumber \\
   &\mp&  \int_{0}^{1/T} d\tau \, \partial_{\tau}\vec x_{i}\cdot (\vec\nabla_{\vec
   x_{i}}\beta) \sin^{2}\frac {k(\beta(\tau)) } 2
    \nonumber \\
   =&\pm&\int_{0}^{1/T} d\tau \, \left( \at^{0}(\tau, \vec x_{i}) + \frac
   {\partial \vec x_{i}}{\partial\tau} \cdot \vec{\tilde a}(\tau, \vec x_{i})
   \right)    \; ,  \label{ber}
\ee
where  the vector potential $\tilde a^\mu$ is given by \pref{atilde}
and the sign depends on whether the spin of the particle is parallel
or antiparallel to $\vec h$.
Note that the angle $\beta(\tau,\vec x_{i}(\tau))$ has both a direct $\tau$
dependence
from the time-dependence of $\mh$, and an indirect one from the
time-dependence of the particle position $\vec x_{i}(\tau)$.

In our particular system where the dynamics fixes both the density and
the magnitude of the polarization (the corresponding modes have large gap
as seen from \pref{modes}) we can calculate the magnetization
energy $E_{M}(\tau)$ in the following approximation,
\be
E_{M}(\vec x) = \mp\half \left[h(\vec x) +
 \half \ell^2 \mh(\vec x) \cdot \vec\nabla^2\vec h(\vec x) + \ldots
\right]
\ \ ,\label{mag}
\ee
where  plus and minus refer to the contributions from particles with spin in
the direction of, or opposite to, the field $\vec h$ respectively.
The first term on the right hand side is unambiguous, it is just the
Zeeman energy, $\mp \half h $, of the particle.
The second term which is $\sim (\vec\nabla\vec S )^{2}$,
requires a more careful treatment involving a discussion of a short distance
effect that goes beyond the arguments used so far: Because of the strong
magnetic field, the electrons will have a rapid cyclotron
motion, in addition to the slow motion of the guiding center. We assume that the
spin, in the adiabatic approximation, will be aligned along the magnetic
field averaged over this rapid motion. (This corresponds to a projection on
the lowest Landau level, as done explicitely in ref. 11.)
We can model this effect by smearing the electron over a distance given by the
magnetic length,
$\ell$,  and evaluate the Zeeman energy by assuming that the spin points in the
direction of, or opposite to, the average $\vec h$,
\be
E_{M}^\pm (\vec x) = \mp\half  \int d^2\delta\, f(\delta) \mh(\vec x)
\cdot\vec h(\vec x + \vec \delta) \ \ \ \ \ . \label{magcorr}
\ee
Taking a gaussian profile, $f(\vec\delta) = \frac 1
{2\pi\ell^2}e^{-\delta^2/2\ell^2}$, for the smeared electron charge, Taylor
expanding $\vec h (\vec x + \vec \delta)$  and performing
the $\delta$ integration, we get \pref{mag}.

Equations \pref{ber} and \pref{mag} allow us to incorporate the effects of
spin in
the path integral in a very simple way: The Berry phase is included by
coupling the particles  to the gauge potential  $\tilde a^\mu$, and $
E_{M}(\vec x)$ is taken outside the path integral in \pref{trace}
since it depends only on $\vec h$.

To proceed, we have to distinguish between partial and full
polarization, and we treat the two cases separately.

\subsection{Partial polarization}

We now return to a second quantized
description and write the trace in
\pref{trace} as a coherent state path integral.
At this point we  also  switch to a bosonic description in terms of a
two-component boson field, $\phi$, and two auxilliary CS-fields
$a_{\alpha}^{\mu}$. After the standard manipulations\cite{zhang}  we get
the following expression for the partition function,
\be
Z[A^{\mu},T] &=& \int {\cal D}\vec h {\cal D}\chi
e^{ -\int_{0}^{1/T} d\tau \int d^{2}x\, (\frac 1 {2V_0}h^{2} + \frac
2 {V_0} \chi^{2} )   } \nonumber \\
&\times& \int {\cal D}\phi{\cal D}\phidag{\cal
D}a_\alpha^\mu e^{-\int
d^{3}x\,{\cal L}[\phi, \nabla^{\mu} \phi, a_\alpha^\mu;A_{\mu}]}
\nonumber  \\
{\cal L} &=& {\cal L}_{0}(\phi, \nabla^{\mu} \phi, a_\alpha^\mu;A_{\mu})
+ E_{M} -(i\chi -\mu)\rho  \label{lag2}
\ee
where
\be
{\cal L}_{0} &=&  \phidag i\nabla_0 \phi -\frac 1 {2m_{e}} |\vec \nabla\phi|^2
   - \frac 1 {2\pi} \ki \alpha \beta
\epsilon_{\mu\nu\sigma}a_\alpha^\mu\partial^\nu a_\beta^\sigma  \\
i\nabla^\mu &=& i\partial^\mu + a_1^\mu + (a_2^\mu +\at^{\mu})
\sigma_{z}  +  A^\mu  \label{cov2}
\ee
and
\be
E_{M} &=& -\,\half \rho \cos\alpha
\left[h +
 \half \ell^2 \mh \cdot \vec\nabla^2 \vec h  + \ldots \right] \ \  .
\ee
The $\pm$ sign in \pref{ber} is here represented by
$\phi^\dagger \sigma_{z}\phi=\rho \cos\alpha$.
This introduces a dependence on the fixed
$z$-direction  (originating from the parametrization \pref{mh} of $\mh$),
and one might think that the global SU(2) spin symmetry is explicitly brooken.
That this is not the case is seen by changing variables to  the field
$\tilde \phi$,
\be
\tilde\phi =  U(\vec k)\phi = e^{\frac i 2 \vec k\cdot\vec\sigma} \phi \ \
\ \ \ ,
\label{newvar}
\ee
which does not change the integration measure, and noting that the
covariant derivative
$\nabla^{\mu}$ transforms like,
\be
U(\vec k)\nabla^{\mu} U^{\dagger}(\vec k) =
D^{\mu} \ \ \ \ \ ,
\ee
where $D^{\mu}$ is the covariant derivative given by \pref{cov},
which depends only on the vector $\mh$.

Finally, substituting \pref{mag} and \pref{newvar} in \pref{lag2},
carrying out the gaussian integrals over the auxiliary
fields $\chi$ and $h$, and dropping the tilde on $\tilde \phi$,
we obtain our model \pref{lag} in the limit of
vanishing Zeeman coupling, and with the potential,
\be
V_{eff} &=& \frac {V_{0}} 8 [\rho^{2}-(\rho\cos\alpha)^{2}] \nonumber \\
&=& \frac {V_{0}} 8 \rho^{2} - \frac {V_{0}} 2 \vec S \cdot\vec S =
\frac {V_{0}} 2 \, \rup\rdown\label{effpot} \ \ \ \ \ ,
\ee
where $\vec S$ is the spin density in \pref{pol}. Note that the
net effect of the above manipulations is to replace the fermionic form
of the spin density operator in \pref{decomp} with the corresponding
bosonic one in \pref{effpot}.
Also note that \pref{effpot} is manifestely invariant under
the gauge and SU(2) transformations discussed in
section 2. The coefficient $\kappa$ is given by
\be
\kappa =   \frac{V_{0}}{2} \rho \ell^2 \cos^2\alpha \label{spinstiff}\ \ \ \ \ ,
\ee
which for $\cos\alpha =1$ agrees with the spin-stiffness
$\rho_s=\kappa/(4\pi\ell^2)$ calculated by Moon \etal for the fully polarized
$\nu=1$ state.

The Zeeman term can be added using the Noether construction as in
section 2, or one can notice that to leading order $\langle \vec s\rangle
= \half \rhobar \cos\ab\, \mh$ so $\mu_e\vec B\cdot \vec S \simeq
\half \rhobar\cos\ab\,
\mu_e \mh\cdot\vec B$, which is just the mean field value of the Zeeman
term in \pref{lag}.

\subsection{Full polarization}

In the case of fully polarized states, the
previous derivation can be considerably simplified.  We again switch to a
second quantized description, but since  all particles have
spin along the direction of $\vec h(\vec x, \tau)$ we can describe the
system with a single scalar field coupled to a single Chern-Simons
gauge potential choosen to change the statistics from fermions to
bosons. The lagrangian corresponding to \pref{cov2} becomes
\be
{\cal L}_{0} &=&  \phidag i\nabla_0 \phi -\frac 1 {2m_{e}} |\vec \nabla\phi|^2
   - \frac l {2\pi} \epsilon_{\mu\nu\sigma}a^\mu\partial^\nu a^\sigma
  \label{lag1f} \\
i\nabla^\mu &=& i\partial^\mu + a^\mu  +\at^{\mu}  +
A^\mu  \ \ \ \ \  ,  \label{cov2f}
\ee
where $\phi$ is a single component field and $l^{-1}$ an odd integer.
Parametrizing $\phi$ as $\phi = \sqrt{\rho}e^{i\vartheta}$, and
performing the same steps as in section 3, \ie fixing unitary gauge,
we arrive at,
\be
{\cal L} &=& \rho(a^0  + \tilde a^0 )
-\frac 1 {8m_{e}}  (\vec\nabla\rho)^2   -
V(\rho)
-\frac \rho {2m_{e}} (\vec a + \vec A +  \atvidv)^2 \nonumber \\
&-&\frac l {2\pi}
\epsilon_{\mu\nu\sigma}a^\mu\partial^\nu
a^\sigma
-\frac {V_0} 2 (\partial _i \vec S)^2  +
\mu_{e}\vec B\cdot\vec S  \ \ \ .
\label{efflagp}
\ee
This lagrangian corresponds to  \pref{efflag} for the case
of partial polarization. A similar treatment of the high frequency mode as
done for partial polarization in section 3 gives the same effective
lagrangian \pref{eff}, with $\alpha=0$. The lagrangian \pref{efflagp} may in
fact be seen as a special case of \pref{efflag} with parameters
$l_{11}=l_{22}=\pm l_{12}$. This is a singular case where the $l$ matrix
is not invertible. One of the Chern-Simons fields act as a multiplier to
enforce the constraint of full polarization, and the $\phi$ field can be
reduced to a single component field coupled to the second Chern-Simons
field.

For a strictly local potential, the resulting potential as calculated from
\pref{effpot} is zero. This just reflects that fermions in a symmetric
spin state do not interact via a local potential. In realistic cases the
potential is of course not strictly local and this will introduce an
effective potential in
\pref{efflagp}.

\section{Concluding remarks }

\renewcommand{\theequation}{7.\arabic{equation}}
\setcounter{equation}{0}
\vskip2mm

We have in this paper presented a mean field model for partially polarized,
as well as fully polarized, quantum Hall states which seems in several
respects preferable to previously used mean field models. It gives the
correct low energy limit, described by a non-linear $\sigma$ model with a
spin wave spectrum determined by the Zeeman splitting. In this limit it
is similar to the low-energy model previously used for the fully
polarized case.  It is of interest to stress some points concerning the
comparison with other mean field models:

For the partially polarized states, there exist already a model
based on a two-component bosonic field $\phi$ which is coupled to two
Chern-Simons fields\cite{hans1}. However, in this model there are no gapless spin waves
(for vanishing Zeeman coupling.) In the present case, which also has a
two-component $\phi$ field and two Chern-Simons fields, such low
energy spin waves exist, due to the presence of the additional variable $\mh$.
The skyrmion solutions are also different in the two models. In the previous
model the skyrmions are small, whereas in the present model the size of the
skyrmions is determined by competition between the Coulomb and Zeeman
terms, and may therefore be large (\ie carry large spin).

For the fully polarized case, there exist a Landau-Ginzburg model
based on a doublet scalar $\phi$ and  a single Chern-Simons gauge
field\cite{lee}. This model has gapless spin waves, but the spin
stiffness depends on the electron mass and not on the potential
energy. That this scale is wrong is a serious problem that has been
pointed out earlier (for example, although Sondhi \etal use the single
CS field model to motivate the $\sigma$-model lagrangian, they use
the phenomenological spin-stiffness in their original skyrmion calculations).
For this reason we believe that our approach is to prefer also for the
fully polarized states.

In addition to the Kohn mode \pref{modes} in the charge density, there is also a
high frequency spin density wave. While the gap of the Kohn mode
follows from a general sum rule argument, this is not the case for the
spin density wave. However, both the high frequency modes are related to
correlations in the ground state wave function, as stressed by Isakov
\cite{isakov}, and the generalized Halperin $(m_{1},m_{2},n)$ states can be
obtained by considering gaussian fluctuations around the ground state mean
field solutions for constant $\mh$.  This makes it plausible that the model
correctly captures also the short distance part of the physics, even though
one needs to go beyond the simple mean field approximation.

The microscopic derivation presented in section 5 gives a rationale for
introducing the new field variable $\mh$. Although the derivation is not
rigorous, in our opinion it captures important elements of the correct
physics. It is also closely related to the more detailed derivation of the
low-energy effective lagrangian given by Moon \etal for the fully
polarized case. At the level of details one should be aware of the following
point: Even if the form of the spin stiffness term \pref{spinstiff} follows
from our general arguments, the numerical coefficient depends on the
precise smearing of the magnetic field due to the cyclotron motion, as well
as on the decomposition of the interaction \pref{decomp} into a spin
dependent and a spin independent part. There is no obvious choice for this
decomposition in our derivation. We have chosen one which gives a bosonic
effective potential  $\sim\rup\rdown$. With this choice, and using
the probability density of a wave packet which is maximally localized
in the lowest Landau level as smearing profile, the spin stifness comes
out with the same numerical factor as in the work by Moon {\em et al.}.

We conclude by two comments on possible extensions of the present work.

1.  The discussion in this paper has been entirely in the context of bosonic
mean field theories. An alternative,
and very successful, approach is the formulation in terms of composite
fermions \cite{jain}.
The CS-mean field description of spin polarized  composite fermions was given
by Lopez and Fradkin\cite{lop1}
and the generalization to partially polarized and unpolarized states by Mandal
and Ravishankar \cite{man}.
Lopez and Fradkin also studied the closely related 2-layer problem\cite{lop2},
and explicit composite fermion
wave functions describing spin waves were studied numerically by Nakajima and
Aoki \cite{aoki}.
The formalism developed in this paper can immediately be carried over to the
fermionic CS-theory, and would
give an alternative description of spin waves in the mean field theory of
composite fermions. A possible
advantage of such a formulation is that it would be manifestly SU(2)
invariant from the outset, and it might
be of interest to compare such a formulation to the ones by Mandal and
Ravishankar, and  by Ray\cite{ray2}.

2. One interesting application of the various effective field theories for the
QH system is to study edge modes.
For sharp edges a dual CS description is very useful, and leads to a
description of the edge modes in terms of chiral
bosons. A dual CS description incorporating spin was given by
Stone,\cite{stone} and recently used by
Milovanovi\'c\cite{milo} to derive the corresponding edge theory.
It might be of interest to derive a dual CS
theory and the  corresponding
edge theory from the CS theory given in this paper. Recently, it has been argued
that, as the confining potential softens, a sharp polarized edge reconstructs by
developing a spin texture \cite{sond,others}.
This phenomenon has been
studied in the bosonic CS theory by Leinaas and Viefers \cite{lein}.
Using the CS theory proposed in
this paper, that analysis could
be extended to partially polarized states, and even in the fully polarized
case, using the lagrangian \pref{efflagp} would
have the advantage of having the correct spin stiffness.

\acknowledgements
We thank Sergei Isakov for describing
his results prior to publication, and  Shivaji Sondhi and Rashmi Ray for
discussions. T.H.H. and A.K. were supported by the Swedish Natural Science
Research Council.

\vskip.3cm
\noindent{email: hansson@physto.se, ak@physto.se, \\
j.m.leinaas@fys.uio.no, unilsson@mercator.math.uwaterloo.ca}

\end{document}